\documentclass[english,aps. pra]{revtex4}
\usepackage[T1]{fontenc}
\usepackage[latin9]{inputenc}
\usepackage{amsmath}
\usepackage{amssymb}
\usepackage{graphicx}

\makeatletter

\newcommand{\lyxdot}{.}

\usepackage{babel}

\makeatother

\usepackage{babel}
\begin{document}

\title{Super-diffusion in optical realizations of Anderson localization}

\author{Yevgeny Krivolapov}

\affiliation{Physics Department, Technion - Israel Institute of Technology, Haifa
32000, Israel.}

\author{Liad Levi}

\affiliation{Physics Department, Technion - Israel Institute of Technology, Haifa
32000, Israel.}

\author{Shmuel Fishman}

\affiliation{Physics Department, Technion - Israel Institute of Technology, Haifa
32000, Israel.}

\author{Mordechai Segev}

\affiliation{Physics Department, Technion - Israel Institute of Technology, Haifa
32000, Israel.}

\author{Michael Wilkinson}

\affiliation{Department of Mathematics and Statistics, The Open University, Walton
Hall, Milton Keynes, MK7 6AA, England}
\begin{abstract}
We discuss the dynamics of particles in one dimension in potentials
that are random both in space and in time. The results are applied
to recent optics experiments on Anderson localization, in which the
transverse spreading of a beam is suppressed by random fluctuations
in the refractive index. If the refractive index fluctuates along
the direction of the paraxial propagation of the beam, the localization
is destroyed. We analyze this broken localization, in terms of the
spectral decomposition of the potential. When the potential has a
discrete spectrum, the spread is controlled by the overlap of Chirikov
resonances in phase space. As the number of Fourier components is
increased, the resonances merge into a continuum, which is described
by a Fokker-Planck equation. We express the diffusion coefficient
in terms of the spectral intensity of the potential. For a general
class of potentials that are commonly used in optics, the solutions
of the Fokker-Planck equation exhibit anomalous diffusion in phase
space, implying that when Anderson localization is broken by temporal
fluctuations of the potential, the result is transport at a rate similar
to a ballistic one or even faster. For a class of potentials which
arise in some existing realizations of Anderson localization atypical
behavior is found.
\end{abstract}
\maketitle

\section{Introduction}

It is well known that a particle moving in a spatially disordered
time-independent potential can exhibit Anderson localization, and
that this is the generic situation in one or two dimensions \cite{Anderson1958,Lee1985}.
However, it is very difficult to see unambiguous evidence for Anderson
localization in solids, because of complications due to electron-electron
interactions. Recently, Anderson localization was demonstrated by
trapping of light propagating in a paraxial disordered optical system
\cite{Schwartz2007}, in the scheme called {}``transverse localization''
\cite{DeRaedt1989}. That experiment had much follow up in optical
setting \cite{Lahini2008,Levi2011}, while related experiments were
carried out also with cold atoms \cite{Sanchez-Palencia2007}. In
the optics experiments in the transverse localization scheme, the
role of time is played by distance along the waveguide, and considerable
care is required to ensure that the disordered modulation in the refractive
index is perfectly uniform along the direction of propagation (which
we take to be the $z$-axis).

If the potential fluctuates in time as well as in space, the arguments
leading to Anderson localization break down. Anderson's original paper
was entitled `The absence of diffusion in certain random lattices',
hence it might naturally be expected that when Anderson localization
is destroyed - it will be replaced by diffusive transport. In the
optical setting of transverse localization, the spatially-disordered
potential corresponds to small random variations in the refractive
index in the $xy$ plane, normal to the propagation direction, whereas
the temporal fluctuations correspond to adding longitudinal fluctuations
(in the $z$ direction) to the disordered refractive index. That is,
the refractive index would have random variations both in the $xy$
plane and in the $z$ direction, but at different rates. In this paper
we consider the effects of such evolving disorder, which as we show,
destroys localization, leading to expansion of the optical beam, by
virtue of evolving disorder. We argue that when Anderson localization
is broken, for a generic realisation of the optical potential, the
result will be super diffusion: expansion of the optical wavepacket
at a rate similar to a ballistic or even faster.

Several different mechanisms for the breakdown of Anderson localization
due to temporal fluctuations of the potential have been discussed,
all of which predict that localisation is replaced by diffusion. Mott
\cite{Mott1979} considered the effect of phonons at low temperatures,
and argued that this gives rise to a diffusive motion of the electrons
termed `variable-range hopping' conductivity. Mott \cite{Mott1969}
also considered the effects of an AC electric field, and suggested
that a resonant interaction dominates the low-frequency response.
It has been argued that an alternative limiting procedure leads to
a distinct type of diffusive response at a low-frequency electric
field, termed adiabatic transport \cite{Wilkinson1991}. Here, we
show that, in contrast to previous work predicting diffusion, the
response to a time-dependent fluctuating spatially disordered potential
may result in super diffusion. Specifically, the diffusion constants
have a sensitive dependence upon energy. Also, if the potential is
time-dependent, the energy of a particle will not be constant. If
the diffusion constant is a rapidly increasing function of energy,
the response to a time-dependent perturbation can be a super-diffusive,
ballistic or even super-ballistic.

Anderson localization is a quantum mechanical effect (at least in
cases where the potential is not large enough to trap the particles
classically). However, as the energy increases the effects of quantum
phenomena decrease, and a classical analysis becomes appropriate.
As we will see in the present work, rapid spreading in configuration
space is related to an increase of the kinetic energy, justifying
the use of the classical (particle) picture. Also, as the energy increases,
the effects of the potential become a weak perturbation. Accordingly,
in order to characterize the asymptotic behavior in the long-time
limit we consider the behavior of a particle moving classically in
a weak disordered potential that is also fluctuating in time.

The random potentials which are prepared in optics \cite{Schwartz2007}
and atom optics \cite{Lye2005,Sanchez-Palencia2007} experiments are
naturally described in terms of Fourier series where the expansion
coefficients are independent random variables. This motivates representing
the random potentials using their spectral content. In practice, there
are a finite number of Fourier coefficients (denoted here by $N$),
but this number may be large. We shall therefore also consider potentials
with a continuous Fourier transform, approximating the limit as $N\to\infty$.

In section \ref{sec:A-particle-in}, we consider the classical dynamics
of a particle in one dimensional potentials that are random in both
space and time, emphasizing the diffusive spread of momentum. In the
case where there is a finite number of Fourier components, the theory
is formulated in terms of Chirikov's resonance overlap criterion \cite{Zaslavskiii1972,Chirikov1979,Lichtenberg2010}.
In the limit as $N\to\infty$, we show how the Chirikov resonances
are related to an expression for a diffusion coefficient $D$ characterizing
random changes in the momentum $p$. In accord with earlier investigations
\cite{Golubovic1991,Rosenbluth1992,Arvedson2006,Bezuglyy2006}, we
conclude that, for generic random potentials the diffusion coefficient
has a universal power-law dependence on momentum $p$, such that $D(p)\sim|p|^{-3}$
as $|p|\to\infty$. This in turn implies that asymptotically in time
the average momentum satisfies $\left\langle p^{2}\right\rangle \sim t^{2/5}$.
The average displacement satisfies $\left\langle x^{2}\right\rangle \sim t^{12/5}$
for one-dimensional systems \cite{Golubovic1991,Rosenbluth1992,Arvedson2006,Bezuglyy2006}
(faster than ballistic) and $\left\langle x^{2}\right\rangle \sim t^{2}$
(ballistic transport on average) for systems with dimension higher
than one \cite{Rosenbluth1992}.

Our discussion in section \ref{sec:A-particle-in} differs from earlier
works analyzing anomalous diffusion in random potentials \cite{Golubovic1991,Rosenbluth1992,Arvedson2006,Bezuglyy2006},
in that we express the diffusion coefficient $D(p)$ in terms of the
spectral intensity of the potential, as well as making the connection
with the Chirikov resonances explicit. This approach highlights some
of the subtleties which can arise when considering real experiments.
In Section \ref{sec:Applications-in-optics} the theory is used to
derive the transport properties for a potential that naturally appears
in optical experiments on Anderson localisation, such as \cite{Schwartz2007}.
We find a family of potentials for which the diffusion coefficient
of the momentum vanishes for high momentum. This suggests that the
diffusive spreading in momentum saturates asymptotically, and therefore
does not exhibit the universal behavior described in previous studies
\cite{Golubovic1991,Rosenbluth1992,Arvedson2006,Bezuglyy2006}. In
the optical experiments such as \cite{Schwartz2007,Lahini2008,Levi2011}
both types of potentials could be readily realized, which allows the
exploration of both transport regimes. The formulae presented here
give a quantitative prediction of the anomalous spread of the beam.
The results are summarized in Section \ref{sec:Summary-and-Discussion}.

\section{\label{sec:A-particle-in}A particle in a quasi-periodic potential}

Some of the experiments which demonstrate an optical realization of
Anderson localization (such as that described by \cite{Schwartz2007})
involve an induction technique, where a change in the refractive index
of a dielectric is induced by an interference pattern generated by
external waves (used strictly to induce the potentials) \cite{Efremidis2002,Fleischer2003}.
Because the optical field defining the disordere is produced by interference
of plane waves, the `potential' is most naturally described in terms
of its Fourier components. For this reason we need to analyze motion
in a quasi-periodic potential.

Consider the motion of a particle of unit mass described by the Hamiltonian,
\begin{equation}
H=\frac{p^{2}}{2}+V(x,t),\label{eq:hamiltonian}
\end{equation}
 where $V(x,t)$ is a one-dimensional quasi-periodic potential of
the form,
\begin{equation}
V(x,t)=\frac{1}{\sqrt{N}}\sum_{m=-N}^{N}A_{m}\exp\left[{\rm i}\left(k_{m}x-\omega_{m}t\right)\right]\label{eq:potential}
\end{equation}
where $A_{-m}=A_{m}^{\ast}$, so that the potential is real. Here
$A_{m}$ are independent (for $m>0$), identically distributed complex
random variables. The expectation values of these variables satisfy
(for $n,m>0$)
\begin{eqnarray}
\left\langle A_{m}\right\rangle  & = & \left\langle A_{m}A_{n}\right\rangle =0\nonumber \\
\left\langle A_{m}A_{n}^{*}\right\rangle  & = & \sigma^{2}\delta_{mn}.\label{eq:A_averages}
\end{eqnarray}
An example of such a variable is $A_{m}=|A_{m}|\exp\left({\rm i}\phi_{m}\right)$,
where $\left|A_{m}\right|$ and $\phi_{m}$ are independent real random
variables, with $\phi_{m}$ uniformly distributed in the interval
$[-\pi,\pi]$. The random variables $k_{m}$ and $\omega_{m}$ are
distributed with the probability density $P(k,\omega)$, which may
be either a continuous spectrum, a sum of delta functions, or a distribution
concentrated on a line in $k$-$\omega$ space. Note that,
\begin{equation}
\left|V(x,t)\right|\leq\sum_{m=-N}^{N}\left|A_{m}\right|/\sqrt{N}\leq2\sqrt{N}\max_{m}\left|A_{m}\right|,
\end{equation}
 however the variance of $\left|V(x,t)\right|$ is equal to
\begin{eqnarray}
\left\langle \left|V(x,t)\right|^{2}\right\rangle  & = & \frac{1}{N}\sum_{m=-N}^{N}\sum_{n=1}^{N}\left\langle A_{m}A_{n}^{\ast}\right\rangle \exp\left[{\rm i}\left((k_{m}-k_{n})x-(\omega_{m}-\omega_{n})t\right)\right]\nonumber \\
 & + & \frac{1}{N}\sum_{m=-N}^{N}\sum_{n=1}^{N}\left\langle A_{m}A_{n}\right\rangle \exp\left[{\rm i}\left((k_{m}+k_{n})x-(\omega_{m}+\omega_{n})t\right)\right]\ .
\end{eqnarray}
Using the assumptions of \eqref{eq:A_averages}, we have
\begin{equation}
\left\langle \left|V(x,t)\right|^{2}\right\rangle =2\sigma^{2}.\label{eq:potential_variance}
\end{equation}
For finite $N$ the potential is a quasi-periodic function of $x$
and $t$, and in the limit of $N\to\infty$, for fixed $x$ and $t$
it is a Gaussian random variable with the variance, $\sigma^{2}$.

The motion of a particle in a potential given by \eqref{eq:potential},
for sufficiently small $|A_{m}|$, (the exact requirement will be
specified below) was analyzed by Chirikov \cite{Chirikov1979}. It
was predicted that the phase-space is built up of chains of non-overlapping
resonances, which are given by the condition,
\begin{equation}
\frac{{\rm d}}{{\rm d}t}\left(k_{m}x-\omega_{m}t\right)=k_{m}p-\omega_{m}=0,
\end{equation}
 which is just the stationary phase requirement. This reduces to the
condition,
\begin{equation}
p_{m}^{\textrm{res}}=\frac{\omega_{m}}{k_{m}}.\label{eq:resonances-1}
\end{equation}
Assuming that the resonances are isolated, starting a particle with
an initial momentum near a resonance, will produce a bounded pendulum-like
motion near that resonance. This can be seen by neglecting all non-resonant
terms in the potential, and making a Galilean transformation to the
frame of reference of the specific resonance. The new Hamiltonian
in this frame of reference is just the time-independent Hamiltonian
of a pendulum,
\begin{equation}
H'=\frac{p^{2}}{2}+\frac{2|A_{m}|}{\sqrt{N}}\cos k_{m}x.
\end{equation}
 We can estimate the width of the resonances, $\Delta_{m}$ (that
is, the range of momentum for which phase points lie on oscillating
trajectories). From energy conservation, $\Delta_{m}^{2}/2=4|A_{m}|/\sqrt{N}$,
\begin{equation}
\Delta_{m}=\sqrt{8|A_{m}|/\sqrt{N}}.
\end{equation}
We now order the resonances, such that, $p_{1}^{\textrm{res}}\leq p_{2}^{\textrm{res}}\leq\cdots\leq p_{N}^{\textrm{res}}$,
and define the distance between the adjacent resonances by, $\delta_{m}=p_{m}^{\textrm{res}}-p_{m-1}^{\textrm{res}}$.
The Chirikov criterion \cite{Zaslavskiii1972,Chirikov1979,Lichtenberg2010}
for trajectories to remain localised close to their initial momentum
is that the resonances do not overlap, that is
\begin{eqnarray}
\left(\Delta_{m}+\Delta_{m-1}\right) & \leq & \delta_{m}\qquad\forall m,\label{eq:non_overlap_cond}
\end{eqnarray}
Under these conditions, the momentum of a particle will not change
appreciably over time. One should note that condition \eqref{eq:non_overlap_cond}
is approximate. To obtain better estimates higher order resonances
should be considered \cite{Lichtenberg2010}.

When the amplitudes of the potential, $\left|A_{m}\right|$, are not
sufficiently small, namely, $\Delta_{m}\sim\delta{}_{m}$, some of
the resonance chains will overlap. It is established \cite{Zaslavskiii1972,Chirikov1979}
that in places where resonances overlap stochastic regions will form,
which will result in a random walk between resonances and therefore
a diffusion in the momentum. In order to observe diffusion the number
of resonances has to be large, since for the diffusion approximation
to be valid a large number of jumps between the resonances has to
occur. We will now obtain the diffusion coefficient, adapting a technique
developed in \cite{Golubovic1991,Rosenbluth1992,Arvedson2006,Bezuglyy2006}
to the case where the potential is described by the statistics of
its Fourier components. In the limit as $N\to\infty$ the Chirikov
resonances become dense in momentum, which appears at first sight
to complicate the problem. However, in this limit the quasi-periodic
potential is replaced by a random potential, and the change of the
momentum in a time interval which is longer than the correlation time
of this potential can be regarded as a stochastic variable. In this
limit, the small changes in momentum which occur over a timescale
which is large compared to the correlation time of the potential can
be treated using a Markovian approximation, which validates the use
of a Fokker-Planck approach.

We will now proceed in line with \cite{Bezuglyy2006}, writing an
expression for the small change in momentum occurring over a time
$\delta t$ which is large compared to the correlation time of the
potential
\begin{equation}
\delta p=\int_{0}^{\delta t}\mathrm{d}t\ F(x(t),t)
\end{equation}
 where $x\left(t\right)$ is the trajectory of the particle and $F(x,t)=\frac{\partial V}{\partial x}(x,t)$
is the force. Defining the force-force correlation function,
\begin{equation}
C(x_{1},t_{1};\, x_{2},t_{2})=\left\langle F(x_{1},t_{1})\, F(x_{2},t_{2})\right\rangle ,\label{eq:12}
\end{equation}
 and assuming that it is stationary, we can express the variance of
the fluctuation of the momentum in the form
\begin{equation}
\left\langle \delta p^{2}\right\rangle =\int_{0}^{\delta t}\mathrm{d}t_{1}\int_{0}^{\delta t}\mathrm{d}t_{2}\, C(x(t_{1})-x(t_{2}),t_{1}-t_{2}),
\end{equation}
and we neglect all the cross-correlations $\langle\delta p_{i}\delta p_{j}\rangle$,
where the indexes $i$ and $j$ correspond to two different intervals.
For this assumption to be true the correlation function $C$ should
decay sufficiently fast, such that for $\tau>\delta t$ it is negligible.
Furthermore, we will expand
\begin{equation}
x(t_{1})-x(t_{2})=p(t_{1}-t_{2})+O(\delta t^{2}),
\end{equation}
which assumes that the force and its time variations are weak enough.
Under these assumptions, we can obtain,
\begin{equation}
\left\langle \delta p^{2}\right\rangle =2D(p)\delta t,
\end{equation}
where the diffusion coefficient is given by,
\begin{equation}
D(p)=\frac{1}{2}\int_{-\infty}^{\infty}C(p\tau,\tau)\,\mathrm{d}\tau.\label{eq:Dif_coef_formula}
\end{equation}
This expression was first obtained by Sturrock \cite{Sturrock1966}.
As a consequence of the increments of momentum being both small and
Markovian, the probability density for the momentum, $\rho(p,t)$,
satisfies a Fokker-Planck equation. Care must be taken over the order
in which derivatives are taken. In \cite{Bezuglyy2006} it is shown
that the correct Fokker-Planck equation is
\begin{equation}
\frac{\partial\rho}{\partial t}=\left(\frac{\partial}{\partial p}D(p)\frac{\partial}{\partial p}\right)\rho.\label{eq:Focker_Planck}
\end{equation}
Using the potential \eqref{eq:potential} we obtain the correlation
function,
\begin{equation}
C(x_{1},t_{1};x_{2},t_{2})=\frac{1}{N}\sum_{m=-N}^{N}\sum_{n=-N}^{N}\left\langle k_{m}k_{n}A_{m}\exp[{\rm i}(k_{m}x_{1}-\omega_{m}t_{1})]\times A_{n}^{\ast}\exp[-{\rm i}(k_{n}x_{2}-\omega_{n}t_{2})]\right\rangle .
\end{equation}
We first perform an average on the $A_{m}$ variables. Using the assumptions
\eqref{eq:A_averages} results in a translationally invariant correlation
function both in space and time,
\begin{eqnarray}
C(x_{1}-x_{2},t_{1}-t_{2}) & = & \frac{\sigma^{2}}{N}\sum_{m=-N}^{N}\left\langle k_{m}^{2}\exp[{\rm i}(k_{m}(x_{1}-x_{2})-\omega_{m}(t_{1}-t_{2}))]+{\rm c.c.}\right\rangle \nonumber \\
 & = & \sigma^{2}\int{\rm d}k\int{\rm d}\omega\, k^{2}P(k,\omega)\left(\exp[{\rm i}(k\left(x_{1}-x_{2})-\omega(t_{1}-t_{2})\right)]+{\rm c.c.}\right),\label{eq:FF_corr}
\end{eqnarray}
where $P(k,\omega)$ is the probability density of $\omega$ and $k$,
which will dubbed in what follows \emph{the spectral content} of the
potential, introduced along with equation (\ref{eq:A_averages}).
Note that the correlation function of the force is a Fourier transform
of $k^{2}\, P(k,\omega)$. To obtain an integrable correlation function,
we therefore have the following requirement on $P(k,\omega)$:
\begin{equation}
\int\mathrm{d}k\mathrm{\int d}\omega\, k^{2}\, P(k,\omega)\equiv M<\infty,\label{eq:P_requirement}
\end{equation}
 which means that $P(k,\omega)$ has a finite support or decays faster
than $k^{-3}$ and $\omega^{-1}$. Note, that distribution $P(k,\omega)$
can contain also atom contributions (delta functions). Equation \eqref{eq:FF_corr}
leads to,
\begin{eqnarray}
D\left(p\right) & = & \frac{\sigma^{2}}{2}\int\mathrm{d}k\int\mathrm{d}\omega\, k^{2}\, P(k,\omega)\int_{-\infty}^{\infty}\mathrm{d}\tau\left(\exp[{\rm i}(kp-\omega)\tau]+{\rm c.c.}\right)\nonumber \\
 & = & 2\pi\sigma^{2}\int\mathrm{d}k\int\mathrm{d}\omega\, k^{2}\, P(k,\omega)\,\delta(\omega-kp).\label{eq:D_p}
\end{eqnarray}
This expression is closely connected to the resonance probability
density, which can be defined as
\begin{eqnarray}
P(p^{\text{res}}) & = & \int\mathrm{d}k\int\mathrm{d}\omega\, P(k,\omega)\delta(p^{\text{res}}-\frac{\omega}{k})\label{eq:res_density}\\
 & = & \int\mathrm{d}k\int\mathrm{d}\omega\,|k|P(k,\omega)\delta(\omega-kp^{\text{res}}).\nonumber
\end{eqnarray}
These two expressions quantify the connection between Chirikov resonances
and phase space diffusion.

Equation \eqref{eq:D_p} can be also interpreted as an integration
of the function $k^{2}\, P(k,\omega)$ over a line with a slope of
$p$, which is just a Radon transform. We can obtain the asymptotic
behavior of \eqref{eq:D_p} for large $p$ following a similar procedure
done in \cite{Arvedson2006,Bezuglyy2006}, by rescaling the variables,
$k'=kp$,
\begin{equation}
D(p)=\frac{2\pi\sigma^{2}}{p^{3}}\int_{-\infty}^{\infty}\mathrm{d}k'\, k'^{2}P(\frac{k'}{p},k').
\end{equation}
Therefore if $P(0,k')\neq0$ in the limit of large $p$ we have,
\begin{equation}
D(p)\sim\frac{D_{0}}{p^{3}},\label{eq: 26}
\end{equation}
 where,
\begin{equation}
D_{0}=\int_{-\infty}^{\infty}\mathrm{d}k'\, k'^{2}\, P(0,k').
\end{equation}
 This scaling of the diffusion coefficient provides an anomalous diffusion
in momentum such that $\langle p^{2}\rangle\sim t^{2/5}$ and $\langle x^{2}\rangle\sim t^{12/5}$
\cite{Golubovic1991,Rosenbluth1992}. The precise prefactors can be
extracted from results in \cite{Arvedson2006} (equation (30))
\begin{equation}
\left\langle p^{2}\right\rangle \sim\frac{5^{4/5}\sin(\pi/5)\Gamma(3/5)\Gamma(4/5)}{\pi}D_{0}^{2/5}t^{\frac{2}{5}},\label{eq:p_2_5}
\end{equation}
 The prefactor in the relation
\begin{equation}
\left\langle x^{2}\right\rangle \sim{\cal C}_{x}t^{\frac{12}{5}}.\label{eq:x_12_5}
\end{equation}
 is given in \cite{Bezuglyy2006} in terms of a one-dimensional integral
(equations (150)-(152)).

In our discussion of optical realisations of Anderson localisation
we will be led to consider potentials for which there is some $p_{\text{max}}$,
such that
\begin{equation}
P(k,\omega)=0\quad:\quad\omega\geq p_{\text{max}}k,\label{eq:potential_cond}
\end{equation}
 namely, the function $P(k,\omega)$ has a non--vanishing support
only outside the wedge with an intercept of $p_{\text{max}}$, than
it follows from \eqref{eq:D_p} that,
\begin{equation}
D(p)=\begin{cases}
4\pi\sigma^{2}\int_{0}^{\infty}\mathrm{d}k\, k^{2}P\left(k,pk\right) & |p|\leq p_{\text{max}}\\
0 & |p|>p_{\text{max}}.
\end{cases}\label{eq:D_p_optics}
\end{equation}
This suggests that under the condition \eqref{eq:potential_cond}
on the potential \eqref{eq:potential} there is a saturation in the
growth of the kinetic energy. Therefore a particle started inside
the part of space with non--vanishing resonance density will not diffuse
to regions of zero resonance density.

To summarise: we have derived the explicit dependence of the diffusion
coefficient on the spectral content of the potential, $P(k,\omega)$,
which is very useful since in many cases in optics and in atom optics
it can be experimentally controlled. An example of this type will
be discussed in the next section.

\section{Applications to optics}

\label{sec:Applications-in-optics}

In this section we will apply the general scheme for the calculation
of the diffusion coefficient to a specific realization of a disordered
potential. In recent experiments examining Anderson localization of
light, the potential was realized by a superposition of plane waves,
which is of a structure similar to \eqref{eq:potential} \cite{Schwartz2007,Levi2011}.
In those experiments light propagates paraxially in a disordered potential:
the signature of localization is that the width of the propagating
beam of light from a coherent and a monochromatic source remains bounded
as it propagates. The disordered potential is produced by utilizing
the photosensitivity of the medium. A powerful polarised writing beam
induces a change in the refractive index, $\Delta n$, of the medium.
The polarization of this beam is selected in such a manner that the
beam does not experience the change in the refractive index that it
induces. The localization experiments are carried out with another
beam (probe) with a different polarization, such that it experiences
the written change in the refractive index
\begin{equation}
\Delta n(\mathbf{r})=B\frac{1}{1+|E_{0}(\mathbf{r})|^{2}/I_{0}},\label{eq: 3.1}
\end{equation}
where $E_{0}(\mathbf{r})$ is the magnitude of the electric field
of the writing beam at position $\mathbf{r}$, and $B$ is a coefficient
proportional to the nonlinear susceptibility of the medium \cite{Schwartz2007,Levi2011}.
$I_{0}$ is some constant background intensity. In this work we will
consider that $|E_{0}(\mathbf{r})|^{2}/I_{0}$ is small (as is in
\cite{Schwartz2007}) so that we can expand,
\begin{equation}
\Delta n(\mathbf{r})=B\left(1-\frac{|E_{0}(\mathbf{r})|^{2}}{I_{0}}\right),\label{eq: 3.1-1}
\end{equation}
furthermore we will assume that the refractive index depends upon
just two coordinates $x$ and $z$, where $z$ measures distance along
the axis of the test beam. We extend the analysis of last section
for a particular potential realization, which is used in many experiments
in optics, showing that when Anderson localization is broken, and
the spread of the beam obeys an anomalous diffusion law. We provide
numerical results to support our conclusions.

The propagation of a monochromatic light beam in a medium with a non-uniform
refractive index is described by the Helmholtz equation for the electric
field $E$,
\begin{equation}
\nabla^{2}E+k^{2}\frac{n^{2}\left(\mathbf{r}\right)}{n_{0}^{2}}E=0,
\end{equation}
where $k=2\pi n_{0}/\lambda$, $\lambda$ is the wavelength of light
in vacuum, $n_{0}$ is the bulk refractive index, and $n(\mathbf{r})=n_{0}+\Delta n(\mathbf{r})$.
Setting,
\begin{equation}
E\left(x,y,z\right)=\psi\left(x,y,z\right)e^{ikz},
\end{equation}
for $\Delta n(\mathbf{r})/n_{0}\ll1$ and when
\[
\left|\frac{\partial^{2}\psi}{\partial z^{2}}\right|\ll2k\left|\frac{\partial\psi}{\partial z}\right|,
\]
the paraxial approximation is invoked. This approximation yields a
Schrödinger like equation for the slowly varying amplitude, $\psi\left(x,y,z\right)$,
\begin{equation}
\frac{i}{k}\partial_{z}\psi=-\frac{1}{2k^{2}}\left(\partial_{xx}+\partial_{yy}\right)\psi-\frac{\Delta n}{n_{0}}\psi,\label{eq:Schrodinger_units}
\end{equation}
 where we have chosen $z$ to be the propagation axis. In this work
we will consider a propagation along a one-dimensional potential,
such that motion is confined to a plane, with the $y$ degree of freedom
frozen.

The fluctuations of the refractive index are achieved by utilizing
the sensitivity of the medium to light at some frequency and polarization,
which allows us to transform a pre-designed interference pattern to
a variation in the refractive index $\Delta n\left(x,z\right)$. The
interference of $N$ plane waves induces a fluctuation $\Delta n(\mathbf{r})$
which is equivalent to a potential of the form of \eqref{eq:potential}.
The total electric field of the writing beam is given by,
\begin{equation}
E_{0}(x,z)=\frac{1}{\sqrt{N}}\sum_{m=-N}^{N}A_{m}\exp[{\rm i}(k_{x,m}x+k_{z,m})z]
\end{equation}
with $A_{-m}=A_{m}^{\ast}$, where $k_{x,m}$ and $k_{z,m}$ are the
$x$ and $z$ components of the wave-number of a plane wave labelled
by an index $m$, for which the magnitude of the wavenumber is $k_{0}$.
The normalization is chosen such that the total power does not change
as a function of $N$,
\begin{equation}
I=\int|E_{0}(x,z)|^{2}\mathrm{d}x\mathrm{d}z=\frac{2V}{N}\sum_{m=1}^{N}|A_{m}|^{2}.
\end{equation}
where $V$ is the volume of the system, does not depend upon $N$.
The resulting change of the refractive index is equal to
\begin{eqnarray}
\Delta n(x,z) & =\frac{B}{I_{0}}|E_{0}(x,z)|^{2} & =\frac{B}{NI_{0}}\sum_{n,m=-N}^{N}A_{m}A_{n}^{*}\exp\left[{\rm i}\left((k_{x,m}-k_{x,n})x+(k_{z,m}-k_{z,n})z\right)\right].\label{eq:delta_n0}
\end{eqnarray}

The experiments are typically performed with the wavevectors of the
driving electric field $E_{0}(\mathbf{r})$ close to the $z$-axis.
This implies that we can use the paraxial approximation for the writing
field $E_{0}(\mathbf{r})$, as well as for the weak probe field. This
justifies the following additional paraxial approximation:
\begin{equation}
k_{z,m}=\sqrt{k_{0}^{2}-k_{x,m}^{2}}\approx k_{0}-\frac{k_{x,m}^{2}}{2k_{0}},
\end{equation}
 to simplify the notation we will set
\begin{equation}
k_{m}\equiv k_{x,m}\ ,\qquad\omega_{m}=\frac{k_{x,m}^{2}}{2k_{0}}.\label{eq:omega_def}
\end{equation}
Then \eqref{eq:delta_n0} simplifies to
\begin{equation}
\Delta n(x,z)=\frac{B}{NI_{0}}\sum_{n,m=-N}^{N}A_{m}A_{n}^{\ast}\exp\left[{\rm i}\left((k_{m}-k_{n})x-(\omega_{m}-\omega_{n})z\right)\right].\label{eq:delta_n_final}
\end{equation}
To simplify comparison with the previous section, we will work in
units where $k_{0}=1$ and $I_{0}=1$, and will designate the paraxial
axis by $t$ (instead of $z$). Using these conventions \eqref{eq:Schrodinger_units}
simplifies to,
\begin{equation}
{\rm i}\partial_{t}E=-\frac{1}{2}\partial_{xx}E+V(x,t)E,\label{eq:Schrodinger_no_units}
\end{equation}
which has the form of the Schrödinger equation in one dimension. In
our model we take the potential to be
\begin{equation}
V(x,t)=\frac{1}{N}\sum_{n,m=1}^{N}A_{m}A_{n}^{\ast}\exp\left[{\rm i}\left((k_{m}-k_{n})x-(\omega_{m}-\omega_{n})t\right)\right],\label{eq:light_potential}
\end{equation}
 where we have absorbed all the constants, including the minus sign
inside the $A_{m}$, and where we have confined to positive indices
because this expression is automatically real. Hence, \eqref{eq:Schrodinger_no_units}
describes a propagation of a paraxial light beam inside a medium with
spatially-varying refractive index.

Having defined the `potential' function for the paraxial equation
by \eqref{eq:light_potential}, we now consider its ray dynamics.
The classical (zero wavelength) system corresponding to \eqref{eq:Schrodinger_no_units}
is a `particle' moving in the potential given by \eqref{eq:light_potential}.
The equations of motion of such `particle' are given by
\begin{equation}
\frac{{\rm d}p}{{\rm d}t}=-\frac{\partial V}{\partial x}(x,t)\ ,\qquad\frac{{\rm d}x}{{\rm d}t}=p,\label{eq:ray_eq}
\end{equation}
where $p$ is the conjugate momentum to $x$. In what follows we will
examine the motion of such a `particle'. We will now proceed with
a similar analysis to the one done in the previous section. Comparing
\eqref{eq:potential} and \eqref{eq:light_potential}, we see that
the resonances of the system are given by
\begin{equation}
\frac{{\rm d}}{{\rm d}t}[(k_{m}-k_{n})x-(\omega_{m}-\omega_{n})t]=(k_{m}-k_{n})p-(\omega_{m}-\omega_{n})=0.
\end{equation}
Or using the definition \eqref{eq:omega_def}
\begin{equation}
p_{nm}^{\textrm{res}}=\frac{\omega_{m}-\omega_{n}}{k_{m}-k_{n}}=\frac{k_{m}+k_{n}}{2}.\label{eq:resonances}
\end{equation}
The number of resonances is $N(N-1)/2$, and their width is,
\begin{equation}
\Delta_{mn}=\sqrt{8|A_{m}A_{n}|/N}.
\end{equation}
 The overlap condition of the resonances is the same as in the last
section.

\begin{figure}
\includegraphics{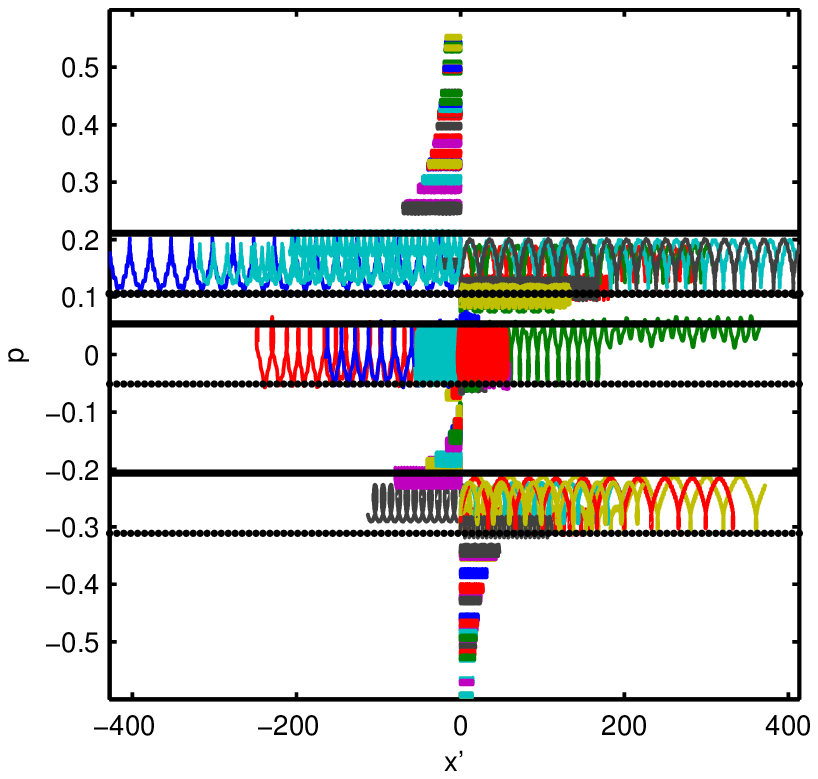}\includegraphics{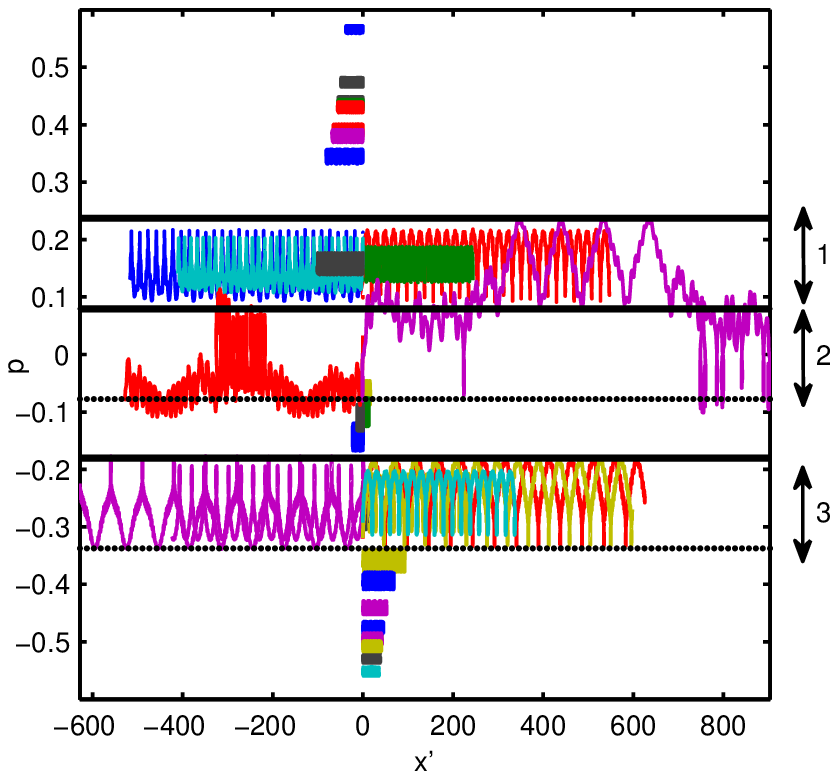}

\caption{\label{fig:3_non_overlap}(color online) A phase-space portrait for
three resonances. Different colors (shades) describe different trajectories.
The horizontal black solid and dashed lines, designate the edges of
the resonance chains. The initial conditions were uniformly distributed
on the $p$ axis $(x(t=0)=0).$ The system was integrated up-to $t=10^{4}$
(dimensionless variables). At the left figure three non--overlapping
resonances are presented, while in the right figure resonances (1)
and (2) overlap.}
\end{figure}

The change of `momentum' (that is, angle) for a ray which propagates
in between the resonances or outside of the region of resonant chains
will average out to zero. This is illustrated in figure \ref{fig:3_non_overlap},
where one can see a transformed phase-space of such a system with
three non-overlapping resonances. The phase-space was sheared by defining
$x'(t)=x(t)-p(t=0)t$. In this manner, trajectories for which the
momentum does not change considerably during the motion of the particle,
are slowly evolving with respect to the transformed $x$ axis, whereas
trajectories which exhibit a large variation in the momentum show
a large excursion in $x'$. From figure \ref{fig:3_non_overlap} we
see that trajectories started within a resonance chain show a large
excursion in $x'$. Nevertheless, all of them are bounded by the boundaries
of the resonance chain. Trajectories started in between the resonance
chains or outside of them have a rather small variation in the momentum,
which decays with increasing the absolute initial momentum.

In figure \ref{fig:3_non_overlap} (right) we illustrate an overlap
between two of three resonances of a system plotted in Fig. \ref{fig:3_non_overlap}
(left). A transition between the overlapping resonance chains is clearly
seen.

As the number of resonances is increased, we can reach a regime where
a large number of these resonances overlap. In this regime we anticipate
that the propagation angle of the ray, $p$, could exhibit anomalous
diffusion. For the regime when all the resonances overlap, we will
compute the diffusion coefficient. The correlation function for the
force is
\begin{eqnarray}
C(x_{1},t_{1};x_{2},t_{2}) & = & \frac{1}{N^{2}}\sum_{n,m=1}^{N}\sum_{i,j=1}^{N}\left\langle A_{m}A_{n}^{\ast}A_{i}^{\ast}A_{j}\right\rangle \nonumber \\
 & \times & \left\langle (k_{m}-k_{n})(k_{i}-k_{j})\exp\left[{\rm i}\left((k_{m}-k_{n})x_{1}-(\omega_{m}-\omega_{n})t_{1}-(k_{i}-k_{j})x_{2}+(\omega_{i}-\omega_{j})t_{2}\right)\right]\right\rangle .\label{eq:light_cor0}
\end{eqnarray}
 Similarly to the previous section we assume
\begin{equation}
\left\langle A_{m}A_{n}^{\ast}A_{i}^{\ast}A_{j}\right\rangle =\sigma^{4}(\delta_{nm}\delta_{ij}+\delta_{mi}\delta_{nj}).
\end{equation}
 This reduces equation \eqref{eq:light_cor0} to
\begin{eqnarray}
C(x_{1}-x_{2},t_{1}-t_{2}) & = & \frac{\sigma^{4}}{N^{2}}\sum_{n,m=1}^{N}\left\langle (k_{m}-k_{n})^{2}\exp\left[{\rm i}\left((k_{m}-k_{n})(x_{1}-x_{2})-(\omega_{m}-\omega_{n})(t_{1}-t_{2})\right)\right]\right\rangle \nonumber \\
 & = & \sigma^{4}\int\mathrm{d}k_{1}\int\mathrm{d}k_{2}\, P(k_{1})P(k_{2})(k_{1}-k_{2})^{2}\nonumber \\
 & \times & \exp\left[{\rm i}\left((k_{1}-k_{2})(x_{1}-x_{2})-\frac{1}{2}(k_{1}^{2}-k_{2}^{2})(t_{1}-t_{2})\right)\right]
\end{eqnarray}
 where $P(k)$ is a density of resonances in $k$ space. We used the
fact that $k$ and $\omega$ are related by a dispersion relation
\eqref{eq:omega_def}. Using the expression for the diffusion coefficient
\eqref{eq:Dif_coef_formula}, we obtain
\begin{eqnarray}
D(p) & = & \frac{1}{2}\int_{-\infty}^{\infty}C(p\tau,\tau)\mathrm{d}\tau=\pi\sigma^{4}\int\mathrm{d}k_{1}\int\mathrm{d}k_{2}\, P(k_{1})P(k_{2})(k_{1}-k_{2})^{2}\delta\left((k_{1}-k_{2})p-\frac{1}{2}(k_{1}^{2}-k_{2}^{2})\right)\nonumber \\
 & = & \pi\sigma^{4}\int\mathrm{d}k_{1}\int\mathrm{d}k_{2}\, P(k_{1})P(k_{2})(k_{1}-k_{2})^{2}\frac{\delta(k_{1}-(2p-k_{2}))}{|k_{2}-p|}\nonumber \\
 & = & 4\pi\sigma^{4}\int\mathrm{d}k_{2}\, P(2p-k_{2})P(k_{2})|k_{2}-p|.\label{eq:D_p_light}
\end{eqnarray}
 Note that if $P(k)$ has a finite support, than $D(p)$ will also
have a finite support. For example for $P(k)=\theta(k_{R}-|k|)/(2k_{R})$,
where $\theta(x)$ is a step function, using \eqref{eq:D_p_light}
we compute the diffusion coefficient
\begin{equation}
D(p)=\begin{cases}
\frac{\pi\sigma^{4}}{k_{R}^{2}}(k_{R}-|p|)^{2} & |p|\leq k_{R}\\
0 & |p|>k_{R}.
\end{cases}\label{eq:D_p_simple}
\end{equation}
Notice that this expression has finite support, so that the asymptotic
relation which leads usually to anomalous diffusion, \eqref{eq: 26},
is not satisfied. In this case a particular form for the potential,
which is used in optical realisations of localisation such as \cite{Schwartz2007},
leads to sub-diffusive growth of $\langle p^{2}\rangle$. For the
case of a large amplitude of the potential, where all resonances overlap,
the variation of the momentum of the trajectories is bounded, in accord
with the prediction that $D(p)$ has finite support.

\begin{figure}
\begin{centering}
\includegraphics{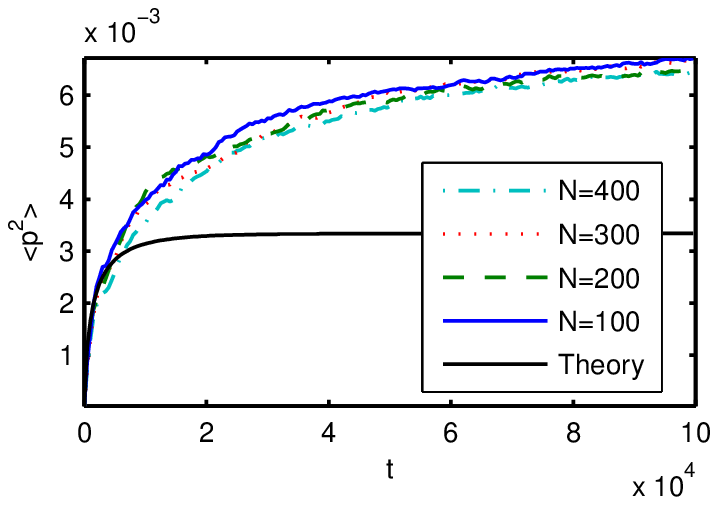}
\includegraphics{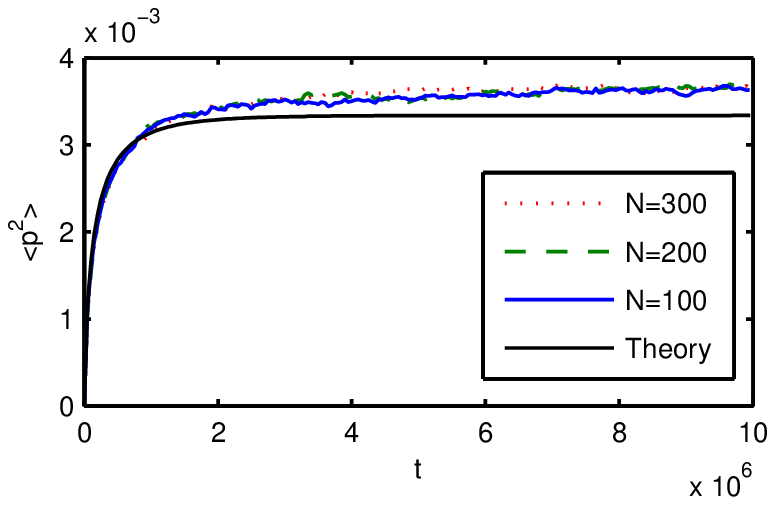}
\par\end{centering}

\caption{\label{fig:diffusion}(color online) $\left\langle p^{2}\right\rangle $
as a function of $t$. Dashed gray lines are numerical solution of
\eqref{eq:ray_eq} for $1000$ initial conditions taken from a Gaussian
distribution. Different shades and styles of lines designate different
$N$ (see legend). The solid think black lines are the numerical solutions
of \eqref{eq:diff_equ} with a diffusion constant given by, \eqref{eq:D_p_simple}.
The parameters are, $k_{R}=0.1$ and $\sigma=10^{-3}$ for the left
figure and $\sigma=10^{-4}$ for the right figure.}
\end{figure}

In figure \ref{fig:diffusion} we compare a direct numerical evaluation
of $\langle p^{2}(t)\rangle$, obtained by averaging solutions of
\eqref{eq:ray_eq} over different initial conditions, with an estimate
obtained from a numerical solution of the Fokker-Planck equation
\begin{equation}
\frac{\partial\rho}{\partial z}=\left(\frac{\partial}{\partial p}D(p)\frac{\partial}{\partial p}\right)\rho,\label{eq:diff_equ}
\end{equation}
 which uses \eqref{eq:D_p_simple} and $k_{R}=0.1$. The classical
trajectories all had initial conditions in the resonance chain. A
good correspondence is found without any fitting parameter.

\section{Summary and Discussion}

\label{sec:Summary-and-Discussion}

We have argued that Anderson localization may not be the absence of
diffusion in disordered potentials, but rather the absence of anomalous
diffusion. More precisely, using the standard argument that at high
energies quantum interference effects are negligible, we used effective
particles to describe the dynamics, and argued that when Anderson
localization is destroyed (due to temporal fluctuations of the potential)
the long-time dynamics is determined by a semi-classical approximation.
For generic potentials, this semi-classical dynamics exhibits anomalous
diffusion in the long-time limit.

Anomalous diffusion in a disordered classical model has previously
been studied in several works \cite{Golubovic1991,Rosenbluth1992,Arvedson2006,Bezuglyy2006}.
This work has considered the issues which arise when considering whether
this type of anomalous diffusion is also relevant to the breakdown
of Anderson localization in optical systems.

The context in which Anderson localization is most readily accessible
to experiment is in propagation in a disordered optical potential
(refractive index) \cite{Schwartz2007} induced by an optical interference
pattern. For this reason, we concentrate upon the case where the potential
is quasi-periodic, resulting from the addition of $N$ waves. We discussed
the influence of the potential on paraxial propagation of a coherent
beam, showing how a semi-classical analysis is related to the Chirikov
resonance overlap criterion \cite{Zaslavskiii1972,Chirikov1979} of
Hamiltonian dynamics.

Gradually increasing the amplitude it is demonstrated how one transforms
from the regime of isolated resonances where no diffusion takes place
(Fig. \ref{fig:3_non_overlap}, left) to a situation where few resonances
overlap (Fig. \ref{fig:3_non_overlap}, right) and spreading that
involves them is found. When the amplitude is even further increased,
such that all the neighboring in momentum resonances overlap and diffusion
in momentum is found (Fig. \ref{fig:diffusion}).

In the limit as the number $N$ of Fourier components increases, the
potential is described in terms of a spectral intensity function $P(k,\omega)$,
and the effect of the potential can be modeled by a stochastic equation,
describing fluctuations of the momentum $p$ with a diffusion coefficient
$D(p)$. We showed how $D(p)$ can be related to the spectral content
$P(k,\omega)$ by \eqref{eq:D_p}, and how this relation can be interpreted
in terms of the Chirikov resonance condition. We also showed that
earlier results on anomalous diffusion, \eqref{eq:p_2_5} and \eqref{eq:x_12_5}
are recovered.

We also considered in some depth the type of potential, such as \eqref{eq:light_potential},
which arises in optical realizations of Anderson localization such
as \cite{Schwartz2007}. We showed that the potential which is used
in these experiments is non-generic, and leads to $\langle p^{2}(t)\rangle$
being practically bounded, rather than exhibiting anomalous diffusion.
This implies that experiments to test the prediction that breaking
localisation leads to anomalous diffusion will have to be carefully
designed, such that the span of the momentum spectrum of the spatial
disorder exceeds the plane-wave spectrum of the initially bounded
probe beam, in order to observe anomalous transport.
\begin{acknowledgments}
We thank Tom Spencer who brought \cite{Rosenbluth1992} to our attention.
We would also like to thank Igor Aleiner, Boris Altshuler and Michael
Berry for informative discussions. This work was partly supported
by the Israel Science Foundation (ISF), by the US-Israel Binational
Science Foundation (BSF), by the Minerva Center of Nonlinear Physics
of Complex Systems, by the Shlomo Kaplansky academic chair, by the
Fund for promotion of research at the Technion. MW thanks the Technion
for their generous hospitality during his visit.
\end{acknowledgments}
\bibliographystyle{apsrev}
\bibliography{QP-potential}

\end{document}